\documentclass{aa}
\usepackage{graphics}
\def\lax    {\ifmmode{_<\atop^{\sim}}\else{${_<\atop^{\sim}}$}\fi}
\def\gax    {\ifmmode{_>\atop^{\sim}}\else{${_>\atop^{\sim}}$}\fi}
\def\kms    {\ifmmode{{\rm ~km~s}^{-1}}\else{~km~s$^{-1}$}\fi}

\def\arcmper  {\ifmmode \rlap.{' }\else $\rlap{.}' $\fi}
\def\arcs   {$^{\prime\prime}$}
\def\arcsper  {\ifmmode \rlap.{'' }\else $\rlap{.}'' $\fi}
\def\arcsgper  {\ifmmode \rlap.^{s }\else $\rlap{.}^s $\fi}

\def\deg      {\ifmmode^\circ\else$^\circ$\fi}     

\def\hper     {\ifmmode \rlap.^{h}\else $\rlap{.}^h$\fi}

\def\m1       {$^{-1}$}

\def\mper     {\ifmmode \buildrel m\over . \else $\buildrel m\over .$\fi}

\def\solmass  {M$_\odot$}
\def\sper     {\ifmmode \rlap.^{s}\else $\rlap{.}^s$\fi}

\def\>           {$>$}
\def\<           {$<$}

\def\simlt       {\lower.5ex\hbox{$\; \buildrel < \over \sim \;$}}
\def\simgt       {\lower.5ex\hbox{$\; \buildrel > \over \sim \;$}}

\def\kms    {\ifmmode{{\rm ~km~s}^{-1}}\else{~km~s$^{-1}$}\fi}

\def\hdos   {\ifmmode {\rm H_{2}}\else ${\rm H_2}$\fi}

\begin{document}

\date{May 2002}
\title{Ripples and tails in the compact group of galaxies Hickson 54}
\thanks{Based on
observations made with the VLA operated by the National Radio
Astronomy Observatory (the National Radio Astronomy Observatory is a
facility of the National Science Foundation operated under cooperative
agreement by Associated Universities, Inc.)
 and on data taken using ALFOSC, which is owned
by the Instituto de Astrof\'{\i}sica de Andaluc\'{\i}a (IAA) and
operated at the Nordic Optical Telescope under agreement between IAA
and the NBIfA of the Astronomical Observatory of Copenhagen.}
 
\author{L. Verdes-Montenegro
\and A. Del Olmo
\and J. I. Iglesias-P\'{a}ramo
\and J. Perea
\and J. M. V\'{\i}lchez
\and M. S. Yun
\and W. K. Huchtmeier}
\institute{Instituto de Astrof\'{\i}sica de
Andaluc\'{\i}a, CSIC, Apdo. Correos 3004, E-18080 Granada, Spain,
lourdes@iaa.es
\and Instituto de Astrof\'{\i}sica de
Andaluc\'{\i}a, CSIC, Apdo. Correos 3004, E-18080 Granada, Spain,
chony@iaa.es
\and Laboratoire d'Astrophysique de Marseille,
Traverse du Siphon - Les Trois Lucs,
13376 Marseille,
France
\and Instituto de Astrof\'{\i}sica de
Andaluc\'{\i}a, CSIC, Apdo. Correos 3004, E-18080 Granada, Spain,
jaime@iaa.es
\and Instituto de Astrof\'{\i}sica de
Andaluc\'{\i}a, CSIC, Apdo. Correos 3004, E-18080 Granada, Spain,
jvm@iaa.es
\and Astronomy Department, University of Massachusetts, Amherst, MA 01003,
USA, myun@astro.umass.edu
\and Max-Planck-Institut f\"{u}r
Radioastronomie, Auf dem H\"{u}gel 69, D-53121 Bonn, 
Germany, huchtmeier@mpifr-bonn.mpg.de}

\titlerunning{Ripples and tails in Hickson 54}
\authorrunning{Verdes-Montenegro et al.}

\maketitle
               
\begin{abstract}
HCG~54 has been classified as a compact galaxy group by Hickson, 
but its nature is uncertain because of its unusual properties.
We present here a study of HCG~54 based on deep optical images and
spectroscopy as well as high spatial and spectral 
resolution HI observations obtained at the VLA. Both optical and radio
data indicate clearly that HCG~54 is a product of a recent
merger involving at least two galaxies. 
Our optical images have revealed a blue elongated structure associated
with HCG~54a embedded in a rounder and redder stellar envelope. 
Several ripples or shells whose center is located 
near HCG~54a and b are also detected. 
These observed features are similar to those found in the
numerical simulations of tidal interactions involving 
two equal mass disk galaxies.   
This system is embedded in an HI cloud 12 kpc in
diameter, and a 20 kpc HI tidal tail emerges from its western edge.
Both the HI distribution and kinematics are consistent with
a recent history of a deeply penetrating interaction or a merger.
Based on the analysis of our new data 
we propose that HCG~54 is the remnant of 
a merger in an advanced stage, whose main body is what has been called 
HCG 54a, while HCG 54b marks the location of a strong starburst 
induced by the interaction, as evidenced by the Wolf Rayet 
stellar population that we detect.
Optical spectra of HCG 54c and d show HII region features and 
are interpreted as tidal debris of the collision undergoing active
star formation.

\end{abstract}

\section{Introduction} 
  
Hickson Compact Groups of galaxies are defined as aggregates 
of four or more galaxies with high galaxy densities similar to
cluster cores (Hickson 1982) and  
low velocity dispersions ($<$ $\sigma$ $>$ = 200
{\rm ~km~s}$^{-1}$, Hickson et al. 1992).  The original catalog, 
composed of 100 groups, was produced without kinematical information,
and was later revised by Hickson et al. (1992) on the basis of 
redshifts obtained for most of the members of the groups. When 
chance projections of pairs and triplets
with discordant galaxies were removed, the catalog 
was reduced to 69 groups. Still, reality of some of the proposed
groups are questioned.  For example, detailed investigations suggest that 
HCG~18 is probably a single knotty galaxy based on the HI and optical 
data (Williams \& Van Gorkom 1988, Plana et al. 2000).

HCG~54 was proposed by Hickson (1989) 
to be a small group composed of four dwarf galaxies, named a through d.
The brightest galaxy has a ``Sdm'' morphological type
while the rest are classified as ``Im''.  Some
authors have later claimed (e.g., Arkhipova et al. 1981)
that HCG 54 is in fact a single galaxy composed of several bright 
HII regions. 
V\'{\i}lchez \& Iglesias-P\'aramo (1998) found  multiple 
emission knots in H$\alpha$. HCG~54 occupies the lower end 
of the  total luminosity, velocity dispersion and optical diameter 
distributions of the HCGs sample, 
with log(L$_B$) = 9.27, log($\sigma_v$) = 2.05 and R$_{\rm 25}$ = 4 kpc. 

We were motivated to study HGC 54 in the context of our 
ongoing studies of HCGs (e.g. Verdes-Montenegro et al. 2001, 
V\'{\i}lchez \& Iglesias-P\'aramo 1998) in order to confirm
or discard its compact group nature based on optical images and spectra, 
together with HI VLA data. 
A large HI tidal tail,
clear signature of a tidal interaction, as well as 
the presence of multiple optical ripples or shells, 
added new motivations to this study.
In particular, the presence of ripples/shells and the apparent Sdm morphology
pose an interesting theoretical problem, since the formation of these
features has been generally attributed to a
bright elliptical accreting a small companion (see 
e.g. Hernquist \& Quinn 1987).
The observations are presented in Sect. 2, the results discussed
in  Sect. 3 and we present possible scenarios in Sect. 4.
A distance of 19.6 Mpc (H$_{\rm 0} = 75 km  s^{-1}Mpc^{-1}$)
is assumed throughout this paper for HCG 54.

\section{Description of the observations}

\subsection{Optical imaging}

We have imaged  HCG 54 in the Johnson R band at the   
2.5m NOT telescope at el Roque de los Muchachos (La Palma)
using the ALFOSC spectrograph during the nights of March 21st and
 27th and May 16th 2001.
The main characteristics of the observations are given
in Table 1. The detector was a Loral/Lesser (CCD\#7) 2048$\times$2048, 
with a spatial scale of 0.189\arcs/pixel, which gives a field of 
about 6\arcmper 5$\times$6\arcmper 5. 
In order to cover a larger field (16\arcsper 6 $\times$ 16\arcsper 7) 
tracing the direction of the HI tail  (see Sect. 3.3), a total of 
10 shifted exposures were taken during the 1st run for 600s each. 
The seeing was varying between 
 1\arcsper 1 and 1\arcsper 4 during the observations.
Ten bias exposures were taken through this run and were used to construct a
median bias  which was subtracted from each image. Pixel-to-pixel variations
were evaluated with 
a median  normalized twilight  flat-field. 
Then  a bidimensional fourth order Legendre polynomial 
was fitted using IMSURFIT
task in IRAF. Images were divided by this flat field. 
After correction of instrumental effects, the atmospheric extinction was 
determined and corrected 
from observations of selected fields from Landolt list of standard
stars  (PG0918 and PG1047). The rms errors of the standard stars in the final
calibration are smaller than 0.07 mag. 
Finally all individual images were combined in a mosaic
by using the  IRAF SQIID package created by K. M. Merrit at NOAO, and 
using SQMOS, XYGET and NIRCOMBINE routines, together with a mask built for each
frame that prevented the mean of the frame edges, as well as bad CCD lines.
The resulting image is shown in Fig. 1, where   
three previously unclassified galaxies A1126+2051, A1127+2054 and 
A1127+2057 are also identified (see Sect. 3.1 and 3.2). 

In the second run,   
a deeper red band exposure was obtained for the central part
in order to better define the faintest optical structures.
The red image is a composition of six 1200s images with the 
Sloan-Gunn $r'$ filter.
A blue image was also obtained in order to be able to derive colour indices
of the different components of the group. 
The Johnson $B$ filter image taken in the 3rd run 
is a composition of four 1200s frames.
Slight dithering was applied between successive frames
for both filters, in order to avoid cosmic rays and bad pixels.
All frames (R, $r'$ and B band images) were finally combined 
in  a search for the best definition of the weakest structures 
(Fig. 2).  None of the nights of these 
 two last runs were photometric, so no calibration was obtained for
the $r'$ and B band images. Instead we adopted the one proposed 
by Hickson et al. (1989) for the $(B - R)$ color map (Fig. 3),
 using
the integrated B-R color of HCG 54b as a reference.

\begin{table}
\caption{Photometric observations.}
\begin{center}
\begin{tabular}{llcl}
 Image& Date&   T$_{ex}$&     Filter\\
\hline
Mosaic& 21/03/2001& 6000&     R-Johnson \\
Center  & 27/03/2001&7200     &     $r'$ Gunn\\
Center  &16/05/2001& 4800    &    B-Johnson\\

\hline
\end{tabular}
\end{center}
\end{table}

\subsection{Optical spectroscopy \label{sec:spectra1}}

The spectra were also obtained with ALFOSC at the NOT 2.5m telescope, with the
same detector used for the images and using  Grism\#4, 7 and 8, 
during April 2000 and March 2001. Table 2 summarizes the long-slit spectra 
taken for this study. The format is as follows: Col. (1) spectrum
identification, (2) direction through which the spectrum was taken, 
(3) date of observation, (4) Grism used, (5) spectral
dispersion, (6) number of exposures, (7) duration of each exposure in
seconds, (8) spectral range, (9) position angle of the slit in 
degrees measured from N to E, and (10) width of the slit.  
The slit widths were chosen to match approximately the seeing of 
the nights. The direction of the slits listed in Table 2
as well as the different zones from which the spectra have been extracted 
along the slits are marked in Fig. 4.
Observations with Grism\#4 were performed in order to derive general 
spectral characteristics and kinematics of  the galaxies, 
and to determine redshifts for possible satellite galaxies in the field.  
Grisms\#7 and \#8 were used to characterize the kinematics with 
a higher spectral resolution than for Grism\#4.
Observations with  Grism\#7  covered with good spectral resolution 
the range including  [OII]3727 and [SII]6717/6731. 
In the observations performed with Grism\#8 the 
 [NII], H$\alpha$ and  [SII] lines were detected and used 
to characterize the kinematics.

The spectra were reduced according to the usual methods, including 
subtraction of a mean bias calculated for each night, division
by a median flat-field obtained for each configuration, as well as wavelength 
calibration based on HeNe calibration lamp spectra obtained after and before 
each exposure. The rms of the bidimensional calibration was 0.4 \AA\ for
Grism\#4 and 0.1 \AA\ for Grism\#7 and Grism\#8. A mean sky was obtained
from object-free sections of each spectrum.  In all cases several exposures 
were taken in order to increase the signal 
to noise ratio and to remove cosmic rays. 
The derived physical parameters described in Sect. 
3.2. are given in Tables 3 and 4.

In order to obtain
redshifts and velocity dispersions we used the cross-correlation
technique derived by Tonry \& Davis (1979). The templates used
for the  spectra showing
emission lines were: the brightest spatial section
of HCG 54a and HCG 54c for the Grism\#8 and Grism\#7 spectra
respectively, together with a synthetic spectrum built from the 
rest frame wavelengths of the emission lines. 
In the case  of the absorption line
spectra, where our goal was to determine the redshift of A1126+2051
and A1127+205, we used as template the spectrum of the radial velocity 
standard giant star HD186176 obtained with the same configuration.
In order to improve the signal to noise ratio of the derived curves
a binning of 2 pixels (0\arcsper 38)
 has been  applied to the spectra sp5 and sp6 
in the spatial direction
and of 3 pixels (0\arcsper 57) to sp1 (Fig. 4). 
The radial velocity data in the direction joining HCG 54a and b are shown in
Fig. 5a, in the HCG 54c - d direction in Fig. 5b
together with the section of sp1 corresponding to HCG 54d, and in the 
HCG 54d - A1126+2051 in Fig. 5c. 
The spectra for HCG 54a and b taken with Grism\#4 are shown in Fig. 6 and
7 respectively.
Spectrophotometric standard stars were observed in order to get a proper 
calibration of the line fluxes in  sp3 and sp4 (Fig. 4).
However, since the nights of April 27 and 28 were non-photometric, only the
relative fluxes of the emission lines are reliable.
The rest of spectra were taken under photometric conditions.

\begin{table*}
\caption{Spectroscopic observations.}
\begin{center}
\begin{tabular}{llcccccccc}
Name& Direction& Date&    Grism&  R(\AA/px)&  N&   T$_{exp}$(s)&   
Wav.-range (\AA)&  PA (deg)&  
Slit width (\arcs)  \\
\hline
sp1& 
HCG 54d -  A1126+2051& 21/03/01&  GR4&  3.0&  3& 1800& 3018-9006&  86&  1.2\\ 
sp2& 
A1127+2057  & 21/03/01&  GR4&    3.0&    2&   1200&    3018-9006& 115&   1.2\\ 
sp3&
HCG 54a - b  & 27/03/01&  GR4&    3.0&    3&   1200&    3018-9006&  65&   1.0\\
sp4&
HCG 54c - d  & 28/03/01&  GR4&    3.0&    3&   1200&    3018-9006&  20&   1.0\\
sp5&
HCG 54a - b  & 27/04/00&  GR8&    1.2&    3&   1800&    5816-8326&  65& 1.2\\ 
sp6& 
HCG 54c - d  & 21/03/01&  GR7&    1.5&    2&   1200&    3815-6815&  15&  1.2\\
\end{tabular}
\end{center}
\end{table*}

\begin{table*}
\caption{Line fluxes relative to H$\beta$
and abundances$^1$ across slit position through HCG 54a - b.}
\begin{center}
\begin{tabular}{lccccccc}
\hline
ION & $\lambda$& $\# 1$ (HCG 54b) & $\# 2 $& $\# 3$ (HCG 54a) 
& $\# 4 $& $\# 5 $& $\# 6 $  \\
\hline
[OII]& 3727& 1.65$\pm$0.05& 4.72$\pm$0.60&4.54$\pm$0.48&4.53$\pm$0.43&
3.86$\pm$0.63 & 5.04$\pm$0.52 \\
$[$NeIII$]$& 3869& 0.35$\pm$0.01& --  &--        & --       &
--             & -- \\
H8+HeI     & 3889& 0.17$\pm$0.01& --  &--      & --      &
--             & -- \\
H$\epsilon$+[NeIII]& 3970& 0.26$\pm$0.01& --   & --     & --    &
-- & -- \\
H$\delta$  & 4100 & 0.24$\pm$0.02& --   & --      & --  &
--  & -- \\
H$\gamma$   & 4340 & 0.45$\pm$0.01& 0.47$\pm$0.04&--  & --  &
-- & 0.43$\pm$0.04 \\
$[$OIII$]$  & 4363& 0.04$\pm$0.01& 0.14$\pm$0.01& --  & -- &
-- & -- \\
HeI      & 4472 & 0.04$\pm$0.01& --     & -- & -- &
-- & -- \\
H$\beta$    & 4861 & 1.00$\pm$0.01&
1.00$\pm$0.05&1.00$\pm$0.02&1.00$\pm$0.02&
1.00$\pm$0.10 & 1.00$\pm$0.03 \\
$[$OIII]   & 4959 & 1.60$\pm$0.02
&0.68$\pm$0.06&0.79$\pm$0.04& 1.56$\pm$0.07 &
0.53$\pm$0.09 & 0.63$\pm$0.04 \\
$[$OIII$]$    & 5007 & 4.83$\pm$0.05
&2.07$\pm$0.15&1.91$\pm$0.08& 3.79$\pm$0.12 &
1.45$\pm$0.24 & 2.02$\pm$0.09 \\
HeI     & 5876 & 0.15$\pm$0.01 & --
&0.25$\pm$0.03& --        & --   & -- \\
$[$OI$]$     & 6300 & 0.05$\pm$0.01 & --    & --
& --     & --     & -- \\
H$\alpha$  & 6563 & 2.68$\pm$0.03 & 3.01$\pm$0.22&
2.97$\pm$0.14& 2.97$\pm$0.17&
2.62$\pm$0.36& 2.81 0.15 \\
$[$NII$]$   & 6584& --  & --
&0.36$\pm$0.02 & --   & --  & -- \\
HeI    & 6678& 0.03$\pm$0.01&
0.26$\pm$0.03& -- & --  & --  & -- \\
$[$SII$]$  & 6717,31& 0.23$\pm$0.01& 0.65$\pm$0.06&
1.03$\pm$0.08& 1.16$\pm$0.06&
0.52$\pm$0.17& 0.93$\pm$0.09 \\
HeI    & 7065& 0.02$\pm$0.01  & --
& --   & --    & -- & -- \\
$[$AIII$]$ & 7136 & 0.08$\pm$0.01  & --
& --    & --       & --   & --  \\
       & & & & & & & \\
1.3*I(6584)/I(3727) & & --     & 0.07    & 0.10  & --  & --  & -- \\
log R$_{\rm 23}$   & & 0.90   & 0.87     & 0.85   &
0.98       & 0.76     & 0.89 \\
P =  $[$(1.3*I(5007)/I(H$\beta$))/R$_{\rm 23}$$]$            && 0.79    & 0.36     & 0.35   &
0.52        & 0.33     & 0.34 \\
log[1.3*I(5007)/I(3727)]    && 0.58    & -0.24     & -0.26   &
0.04       & -0.31     & -0.28 \\
t$[$OIII$]$ (10$^{4}$ K)  & & 1.05  & --     & --
& --     & --  & -- \\
t$[$OII$]$  (10$^{4}$ K)  & & 1.14  & --     & --
& --    & --  & -- \\
& & & & & & & \\
10$^4$ O$^{2+}$/H$^+$& & 1.46   & --     & --    &
--     & --  & --  \\
10$^4$ O$^+$/H$^+$   & & 0.36   & --     & --    &
--     & --  & -- \\
12 + Log O/H         & & 8.26    & 8.30     & 8.30   &
8.20     & 8.10:,8.60: & 8.25:  \\
LogNe$^{2+}$/O$^{2+}$& & -0.67  & --   & --  & --  & --  & -- \\
\hline
\end{tabular}
$^1$ Uncertain values are indicated with ``:''.
\end{center}
\end{table*}

\begin{table*}
\caption{Line fluxes and abundances across slit position through HCG 54c
- d}
\begin{center}
\begin{tabular}{lcccc}
\hline
  & $\lambda$  & $\# 1 $  & $\# 2 $  & $\# 3$  \\
\hline
        & & & & \\
$[$OII$]$    & 3727 & 3.26$\pm$0.37 & 4.53$\pm$0.58 &
3.35$\pm$0.50  \\
H$\gamma$   & 4340 & 0.46$\pm$0.04 & --        &
--       \\
H$\beta$  & 4861 & 1.00$\pm$0.02 & 1.00$\pm$0.08 &
1.00$\pm$0.05 \\
$[$OIII$]$   & 4959 & 0.76$\pm$0.05 & 1.64$\pm$0.16 &
0.80$\pm$0.08 \\
$[$OIII$]$    & 5007 & 2.57$\pm$0.12 & 4.04$\pm$0.38 &
1.98$\pm$0.12 \\
H$\alpha$  & 6563 & 3.12$\pm$0.18 & 2.97$\pm$0.36 &
2.69$\pm$0.21 \\
$[$SII$]$  & 6717,31 & 0.78$\pm$0.17 & 0.91$\pm$0.17 &
0.79$\pm$0.10 \\
       & & & & \\
log R$_{\rm 23}$& & 0.82  & 0.99  & 0.77  \\
P  = $[$(1.3*I(5007)/I(H$\beta$))/R$_{\rm 23}$$]$& & 0.51 &0.54& 0.43 \\
log[1.3*I(5007)/I(3727)]& & 0.01      & 0.07    & -0.12\\
12 + Log O/H  & & 8.10:, 8.40:  & 8.20  & 8.10:, 8.45: \\
\hline
\end{tabular}
\begin{list}{}{}
\item[$^{\rm 1}$] Uncertain values are indicated with ``:''.
\end{list}
\end{center}
\end{table*}

\subsection{VLA HI Observations}

We have mapped HCG 54 with the VLA in the C array in August 1997.
The synthesized beam is 20\arcs\ $\times$ 16\arcsec\ (1.9 $\times$ 
1.5 kpc at a distance of 19.6 Mpc) after natural weighting of the data. 
 A velocity range between 1085 \kms\ and
1730 \kms\ was covered with a velocity resolution of 10.4 \kms .  
The rms noise is 0.33 mJy/beam corresponding to a column
density of 1.2 $\times$ $10^{19}$ at cm$^{-2}$.    
 Assuming  an HI linewidth of 30\kms, the achieved HI
mass detection limit is about 10$^6$ M$_{\odot}$\footnote{    
Computed as $M(HI) = 2.36 \times 10^5~D^2 S \Delta V$,
where $D$ is the distance in Mpc and $S \Delta V$ is the velocity
integrated HI flux in Jy {\rm ~km~s}$^{-1}$.}.

We have detected emission above 3$\sigma$ in the velocity range 
1334.2 to 1490.2\kms . The integrated emission is presented
in Fig. 8a  superposed on the R-band image,
and the velocity field is presented in Fig. 8b.  The velocity
channel maps are shown in Fig.~9.
The total spectrum has been obtained by integrating the emission in the
individual channel maps  (Fig.~10, solid line). 
The total HI line flux
detected, 6.23  Jy {\rm ~km~s}$^{-1}$, is in good agreement
with the single dish measurement by Huchtmeier (1997; 
Fig. 10, dotted line). The total  
HI mass derived is 5.6 $\times$ 10$^8$ \solmass.

\section{Results}

\subsection{Optical images}

The brightest central part of HCG 54 (Fig. 2) shows a dominant knot (HCG 54a) 
embedded in an elongated feature visible in the blue and red bands
at a PA of 65$^{\circ}$. It is more extended
to the SW, where a very compact bright knot is located (HCG 54b) which is even
more prominent in H$\alpha$ (see V\'{\i}lchez \& Iglesias-P\'aramo 1998). 
The elongated area defined by HCG 54a and b is connected to the NE with 
the weaker but still prominent knots HCG 54c and d.

Large variations in the B-R color index
are apparent throughout the whole system (from 0.15  in HCG 54d
to 0.70 around HCG 54a, Fig. 3). 
The elongated structure centered in HCG 54a has a blue color (0.5), 
likely a signature of a
recent star forming episode, and is embedded in a rounder region
that shows the reddest colours in HCG 54 (B-R = 0.7). 

The B-R color index for HCG 54b is
surprisingly red for a strong star formation burst like the one hosted by this
galaxy. However, we argue that this is due to the contamination by the H{$\beta$},
[O{\sc iii}] and H$\alpha$ emission lines. After computing the equivalent
widths of these lines (EW(H$\beta$) = 155 \AA , EW([OIII]4959) = 246 \AA ,
EW([OIII]5007) = 728 \AA , EW(H$\alpha$ + [NII]) = 798 \AA ),
we estimated that the corrected B-R color should be
0.46 magnitudes bluer than the one measured directly from the color map. Thus,
we would obtain B-R = 0.20, which is more consistent with a very young stellar
population.

Three shell-like structures are found (noted as t1 to t3 in Fig. 2),
centered roughly on HCG~54a and b.
A fourth optical feature is located to the SW and marked as t4.
We detect all of these features in a $R$-band surface brightness range of 
24 to 27 mag/(\arcs )$^2$. 
The outer shells  (t2 and t3) although redder than the internal knots, 
have blue colours ($\sim$ 0.5) typical of irregular galaxies
(see Fukugita et al. 1995). 
The inner shell (t1) is slightly redder (0.60), probably due to contamination
from the central parts of HCG 54. Numerous unresolved knots are detected
everywhere in the area, such as the one marked in  Fig. 2 with a ``k''.  
No stellar counterpart to the HI long tail is detected down to 
27  mag/(\arcs )$^2$ in R.

We found three faint galaxies in our large R-band frame (Fig. 1),
that are named A1126+2051, A1127+2057 and A1127+2054. The first 
two are background galaxies (Sect. 3.2) 
while the last one is at the velocity of the system according 
to the HI data (Sect. 3.3).  The optical image of this galaxy
shown in Fig.~11 reveals that the inner and the outer
isophotes are off-centered.  When measured at an R 
isophote of 26 mag/(\arcs )$^2$ the optical size is 2.6 kpc
and the total magnitude in R is M$_R$ = -14.2 mag. 
The surface brightness derived as a function of radius is exponential,
consistent with that of a disk system (see Fig. 11b).

\subsection{Description of the Spectra \label{sec:spectra}}

From the Grism\#4 spectra sp1 (Fig. 4) and sp2 (Table 2) 
we have determined that A1126+2051 
and A1127+2057 are not at the group redshift.
A1126+2051 shows absorption lines of CaII, G band,
  Mg I and Na lines and has a redshift of z = 0.15 (see also Sect. 3.3).
A1127+2057 also shows an absorption spectrum typical of early type galaxies
and gas at a redshift of  z = 0.052.

We have studied  the kinematics of the central parts of the group with the 
spectra obtained with  Grism\#7 and \#8. The spectrum along the HCG 
54a - b direction (sp5, Fig. 4)
shows emission lines along the 249 spatial sections (47\arcs ), and the 
obtained velocity curve is plotted in Fig. 5a, where the zero
position corresponds to the center of HCG 54b. 
We mark in the figure the continuum extent of HCG 54a and mark with an arrow
the position of the emission line maxima. 
The spatial sections closer to HCG 54a (5\arcs\ diameter) 
traces a distorted rotation curve with an amplitude of 
45\kms. At both sides of the center of HCG 54a we find discontinuities
in the velocity that correspond  in the R band to distorted isophotes.
The sections in the direction of HCG 54c (R1 in Fig. 5a) have
three well differentiated knots with emission lines, whose left end
is the external part of HCG 54c. The strongest knot presents a 
velocity gradient of 50\kms\ in 4\arcs\ that corresponds to the 
contact area between HCG 54a and 
HCG 54c (R3).  The velocities in the direction of HCG 54b 
show irregular changes of up to 70\kms, as is the case of 
R2 (Fig. 5a) that appears to be an HII region detached from the general trend 
of the velocity curve.  The area of HCG 54b has a peculiar 
kinematics (Fig. 5a) that could be consistent
with infalling gas, in the case that  the background emission is hidden 
by extinction so that we are only observing the motions of the 
 foreground component.

In Fig. 5b we show the velocities in the slit through
HCG 54c - d (sp6, Fig. 4), and the regions corresponding to 
HCG 54c and d in the R image are also marked. 
These regions have a weak continuum, and 
none of them show a rotation curve. HCG 54d has irregular motions,
while HCG 54c shows a nearly constant velocity following a U-shape with
an amplitude 30\kms. 
We do not detect continuum emission toward 
the third emitting region in the direction of HCG 54a (Fig. 5b), while
the velocities increase continuously until it reaches the values
characteristic of HCG 54a.
Finally we find a region  with a decoupled velocity  (R4)
that is located between HCG 54c and d.

In the lower resolution spectrum crossing HCG 54d (sp1, Fig. 4) 
we detect a velocity gradient of only $\sim$ 70 \kms\ within 7\arcs.
In Fig. 5c we show the sections where we have detected emission,
where the center corresponds to the continuum peak of HCG 54d along
the slit.
The presence of several knots in our R band image along the 
slit direction suggests that  the continuity of the observed
velocity gradient might be due to smearing of individual components.

We have extracted individual spectra for 9 zones from the data taken
with Grism\#4 (sp3 and sp4, Fig. 4). Six of them are along the slit
position joining HCG 54a -b and three along the HCG 54c - d direction
(Fig. 4).  The spectrum of HCG 54a (sp3, Fig. 4) shows strong Balmer
absorption lines as well as lines of CaII, Gband, and possibly MgI, 
while Mg$_{\rm 2}$ is barely detected. The spectrum shows a blue continuum, 
characteristic of a post-starburst
population (Fig. 6).
The spectrum corresponding to knot b is remarkable (Fig. 7), with a
high excitation and the presence of WR features at $\sim 4650 $ \AA \
over a flat continuum, indicative of a young and strong burst of star
formation.  Fig. 7a shows the full spectrum of HCG 54b, while
Fig. 7b shows a detail of the spectrum around the Wolf-Rayet (WR)
feature.  The temperature sensitive [OIII] line at $\lambda$ 4363 \AA
\ was measured in this spectrum, giving a t([OIII]) temperature of
10485 K, implying an oxygen abundance of 12+LogO/H = 8.26 (see Table
3).  According to Schaerer \& Vacca (1998), the measured equivalent
width of the WR $\lambda$ 4650 \AA \ feature in knot b, EW(WR) = 9
$\pm$0.9 \AA , implies an age of the burst between 3 and 4 Myr.

For 
the zones showing emission line spectra (sp3, sp4, Fig. 4) we present in 
Tables 3 and 4 their fluxes relative to H$\beta$ as well as the derived 
physical conditions of the ionized gas.
  Reddening corrected line fluxes relative to H$\beta$ are presented
for the 9 individual spectra extracted along slit positions sp3 and
sp4 (Table 2, Fig. 4).
For each zone given in Tables 3 and 4 the following ionization
structure parameters have been derived in order to perform the
abundance analysis (as detailed below): R$_{\rm 23}$, denotes the
abundance parameter after Pagel et al. (1980) which is defined as
[I(3727) + I(5007) + I(4959) ]/I(H$\beta$), P denotes the abundance
parameter defined by Pilyugin (2000), as quoted in Tables 3 and 4, P =
[1.3 $\times$ I(5007)/I(H$\beta$)]/R$_{\rm 23}$.
The excitation, defined as  1.3  $\times$ I(5007)/I(3727), and
the nitrogen to oxygen abundance indicator, 1.3  $\times$ I(6584)/I(3727),
are also quoted in Table 2. Note the small range of variation of
R$_{\rm 23}$ in contrast with the large  variations shown by the
excitation along the slit.  The electron temperatures t[OIII]
and t[OII],  corresponding to the ionization zones of [OIII] and
[OII] respectively, have been derived for region $\#$1 and listed
in Table 3.
For this region the ionic and total abundances of Oxygen,
O$^{2+}$/H$^{+}$, O$^{+}$/H$^{+}$ and O/H respectively, as well as the
abundance ratio of neon to oxygen, Ne$^{2+}$/O$^{2+}$, have been
calculated and are presented in the table. For the rest of the studied
regions, only the total abundance of oxygen has been estimated (see
Tables 3 and 4).  

Given that the temperature sensitive line
[OIII]lambda 4363A was measured only for knot b, we have to rely on the
empirical calibration in order to derive their abundances. We have used the
calibration of the oxygen abundance  versus R$_{\rm 23}$ (cf. Pagel et
al. 1980) as parameterized by McGaugh in Kobulnicky et al. (1999), and
following the P-method (Pilyugin 2000), in order to provide an
estimation of 12+Log O/H.
 The [NII]/[OII] line ratio, when observed,
was used to discriminate between the lower and upper branch, though often
the 6584 line was severely blended with H$\alpha$ and was not
measured. For those zones with R$_{\rm 23}$ $\ge$ 0.9 we have assumed an
average abundance of 12+logO/H = 8.2 as indicated by the calibration.
All the zones selected are consistent with 12+Log O/H = 8.2, within the
errors of the empirical calibration. This abundance is typical of galaxies
like the Large Magellanic Cloud and the outer disks of late type spirals.

\subsection{Neutral Hydrogen \label{sec:HI}}

The integrated emission of neutral hydrogen (Fig. 8a) shows a 
NE-SW distribution with extensions to the SE and SW, and 
a long tail with a projected size of 20 kpc to 
the NE. The velocity field is quite perturbed, but still shows 
a velocity gradient similar to a rotating disk  with a  twisted major axis
(Fig. 8b). This reflects itself also in the asymmetry of the HI 
line integrated profile (Fig. 10). 

The situation is more complex when the channel maps (Fig. 9) 
are examined.  The sizes of the optical knots are small compared
with the VLA synthesized beam, and tracing the HI kinematics
with respect to the optical features is difficult.  Nevertheless
these channel maps reveal the details that are lost in the
integrated emission image shown in Fig.~8a.
Except for the large HI tail to the northeast, most of the HI
emission arises within the faint optical extent of HCG~54.
Bright HI features directly associated with the bright optical
ridge of emission are seen in the channel maps with velocity
range between 1365 - 1490 km s$^{-1}$.  The overall
velocity field is along the length of the bright optical
ridge in a manner consistent with that of rotation, but 
clear evidence for a velocity gradient in the perpendicular
direction is also present, increasing in velocity from NW to SE.
Since stars and gas inside the tidal radius are generally 
unaffected by a tidal interaction, the observed kinematic 
disturbance suggests an involvement of a deeply penetrating 
interaction or a merger.  

Most of the high surface brightness HI features associated
with the fainter outer optical envelope occur on the west side of
the optical galaxy, closely associated with the optical tidal
features t1, t2, and t4 (see Fig.~\ref{fig:B+R}).  
Both the Y-shaped HI morphology
in Fig.~8a and the shifting of the focus of the iso-velocity
contours to the west of the optical peaks in Fig.~8b are direct
results of the large amount of HI associated with these tidal
features.  This one-sided appearance may indicate that only
one of the progenitor systems was HI-rich if HCG~54 is mainly
a product of a merger involving two late type galaxies -- i.e.
the progenitor responsible for the tidal feature t3 had
relatively little HI associated with its stellar disk.
The HI extension to the southwest, closely associated with
the optical feature t4, occurs mainly in the velocity ranges
of 1334 to 1428 \kms\ and contains 1.4 $\times 10^8 M_\odot$
of HI.  The long HI tail associated with either t1 and/or t2
occurs in the velocity range of 1376 to 1459 \kms\ and 
contains about 10$^8 M_\odot$ of HI.  Combined together,
these two outer HI features account for more than 40\% of the
total HI detected in this system.

Several HI clumps that are detached from the main body of HI are 
also seen at 3-5 $\sigma$ levels in these channel maps, and they
indicate a more extensive debris field associated with this
system.  A velocity coherent string of HI clumps forming a nearly
complete loop or a ring is seen in the channel maps within the velocity
range of 1438 to 1469 \kms, largely to the northeast.  Some of these
features make up part of the 20 kpc long HI tail to the north,
but their appearance in the channel maps, particularly at 1438 \kms\
and 1448 \kms, suggest a more ring-like morphology, similar to
those seen in collisional ring galaxies (e.g. Higdon 1996).  
The stream of HI clumps extending to the west of
the main body seen in Fig.~8a form a second distinct, velocity coherent
structure appearing at velocities of 1386-1407 \kms, and these
channel maps suggest its origin being more to the south of the
main body rather than a linear east-west structure.  The total HI
mass associated with this feature is about  3 $\times 10^7 M_\odot$. 

The overall velocity field of the bulk of HI 
delineates an elongated structure with an axis ratio that 
would correspond, if intrinsically circular, with an  approximate 
inclination of 50$^{\circ}$ and a position angle of 70$^{\circ}$, similar to 
the one traced by the direction of HCG 54a - b (PA=65$^{\circ}$). 
The amplitude derived from the velocity field (140\kms ) and 
deprojected according to this inclination gives an overall velocity
gradient of $\sim$ 183\kms.  If this is interpreted as a Keplerian 
rotation, we estimate a dynamical mass for the
system of M$_{tot} \sim 10^{10} M_{\odot}$.

A dwarf companion galaxy A1127+2054 is found at a projected distance
of 27 kpc northeast of HCG~54a (see
Sect. 3.1., Figs. 1 and 11).  Its location also corresponds to
a distance of 8 kpc from the tip of the 20 kpc long HI tail,
in the same projected direction.  The HI channel maps show associated
emission in 3 channels covering velocities of 1386-1407 \kms.
A close examination suggests a 
central depression in the integrated emission (see Fig. 11c).
The velocity field  is consistent with a slowly rotating disk with 
a velocity amplitude of 30 \kms\ (Fig. 11d).
The HI extent at 3$\times$ 10$^{19}$ at cm$^{-2}$ (3$\sigma$ level)
is 45\arcs\ $\times$ 25\arcs\ (4.3 kpc $\times$ 2.4 kpc) and contains 
1.9 $\times$ 10$^7$ M$_{\odot}$ of HI. The atomic component seems 
to be perturbed since it is not well centered on the optical component.

In the superposition of the integrated HI emission over the deep optical
image shown in Fig.~\ref{fig:HImoments}a, the HI isolated peak located
about 2\arcmin\ east of HCG~54a appears to have an optical counterpart
in a bright, compact galaxy.  From the optical spectrum we obtain,
however, that this is a chance superposition as this galaxy A1126+2051
is found to be a background object with $z=0.15$ 
(see Sect.~\ref{sec:spectra}).
At the position of this galaxy we find 
HI emission in two channel maps, with no signature of systematic motions. 
Since the gas is located  close to the center of the group, it is 
possibly part of the tidal debris. This is supported when a 
sharpening of the  optical image
of A1126+2051 is performed, suggesting the existence of two objects,
a symmetric disk structure whose peak is located in the slit,
and then corresponds probably to the background galaxy, plus a 
 blob that might be a stellar counterpart of the HI emission.

\subsection{Radio Continuum Emission}

Using the data from the line-free channels, a 1.4 GHz continuum
image was constructed.  No significant continuum emission 
associated with the optical galaxy is detected at the $3\sigma$
limit of about 0.5 mJy beam$^{-1}$.  The New VLA Sky Survey
image also gives a comparable upper limit of about 1.5 mJy
for about a 3 times larger beam.  HCG~54 was detected by IRAS
in the 12 $\mu$m, 60 $\mu$m, and 100 $\mu$m bands with flux densities of
0.24, 0.50, and 0.84 Jy, respectively, with the derived FIR luminosity 
of $L_{FIR}=3\times 10^8 L_\odot$ (Verdes-Montenegro et al. 1998).
The upper limit of 1.5 mJy for 1.4 GHz radio continuum makes
this object slightly underluminous in the radio with a lower limit on the 
q-value of about 2.7 (see Yun et al. 2001).
The enhanced IR luminosity resulting from the relatively high 
dust temperature, inferred from the FIR 
($S_{\rm 60}/S_{\rm 100}=0.60$) and mid-IR ($S_{\rm 25}/S_{\rm 60}=0.48$) color,
offers a natural explanation for the relatively weak radio continuum.
These warm infrared colors are also consistent with the evidence for 
significant WR  activity detected in the optical spectra 
(see Sect.~\ref{sec:spectra1}).

\section{Discussion}

We have identified clear signs of tidal interactions and 
possibly a merger toward HCG~54.  The morphology and kinematics of the 
atomic gas, a sensible tracer of interactions, is strongly 
perturbed, including a long (20 kpc) HI tidal tail, plus 
two more HI extensions that emerge from the main body.
The stellar component also shows features that are characteristic 
of interactions or mergers, in particular several  shells around 
the bright central area.  We propose here that the formation of 
the HI tail and the optical ripples involved interactions and
a merger of at least two galaxies.

\subsection{The Isolated Environment of HCG 54: only one (dwarf) companion}

The proximity of HCG 54 to our Local Universe (v = 1470\kms ) 
makes the study of its large scale environment difficult. 
We find in the NASA Extragalactic Database (NED)\footnote{This research 
has made use of the NASA/IPAC Extragalactic Database (NED) which is 
operated by
         the Jet Propulsion Laboratory, California Institute of Technology, 
under contract with the National
         Aeronautics and Space Administration.} 
that  all galaxies with measured
redshifts in a square of 1 Mpc radius centered in HCG 54
have velocities larger than 4000\kms.
Within a distance of 500 kpc none of
the galaxies without redshift determination have a similar size or 
magnitude as the HCG 54 members. In fact,
the closest and brightest galaxy to HCG 54, A1127+2057,
is a background galaxy (see Sect. 3).  Similarly, another
optically selected potential neighbor A1126+2051 is also
shown to be a background galaxy.

The VLA primary beam (30\arcmin) covers a diameter of 170 kpc
at the distance of HCG~54, and we found only one object 
(A1127+2054, see Sect. 3) in this area at the group redshift
in HI emission, within the observed range of 645 \kms. 
The dwarf disk galaxy A1127+2054 is the only identified companion in 
the environment of  HCG 54 and is located at 20 kpc from its center.

\subsection{Origin of the stellar ripples/shells \label{sec:ripples}}

Shells are usually thought to form in unequal mass interactions 
involving a bright elliptical accreting a small companion 
(see  e.g. Quinn 1984, Hernquist \& Quinn 1987). 
The work by Schweizer \& Seitzer (1988) showed that 
a smooth triaxial potential, or even a disk, 
can also give rise to the formation of 
shells and ripples in the stellar component of 
an interacting smaller companion.
It does not seem to be the case in HCG 54, 
where we do not detect a large bulge component nor any other massive system. 
The two galaxies should have been most probably late-type systems, judging
from the post-starburst nature of the spectrum shown by HCG 54a
(without clearly detected features typical of early type systems) which
could be  invoked as the possible nucleus of the system. 
Furthermore, the low chemical abundance 
derived from the ionized gas, 12+log(O/H) $\sim$ 8.3, is more typical
of irregular galaxies. 
Few examples of shells around 
low mass  late type galaxies are found in the literature, as NGC 7673 
(Homeier \& Gallagher 1999). If we relax the low mass 
criteria there is also NGC 3310 (see Mulder \& van Driel,
1996) which is a peculiar Sbc. In both cases a tidal interaction origin 
has been attributed  to the shells.
In general, little observational evidence exists for 
 ripples in disk-disk interactions,  (see e.g. Homeier \& Gallagher
1999, Charmandaris \& Appleton 1996, Kemp \& Meaburn 1993 or 
Schweizer \& Seitzer 1988).
Probably for this reason few models are found focused on reproducing
HCG 54-like systems. In this sense, the simulations presented by 
Hernquist \& Spergel (1992) seem to provide with a more suitable model
to describe the system. The model represents 
a merger of two equal disks galaxies in a close collision
from a parabolic orbit. At the final stages the remnant has
a central bar-like structure  embedded in a disk like envelope, 
together with several shells at different radii.
This resembles quite well the 
optical structure of HCG 54, centered on HCG 54a,
with an associated bar-like bluer structure, whose colours are suggestive 
of star formation,  
in the middle of  a red disk-like component and surrounded by 3 shells.
The  shells are centered somewhere close to HCG 54 a and b, 
but the data do not allow a  determination of  the precise location of 
their focus (Sect. 3.1). The outer shells (t1 \& t3, see Fig. 2) 
have bluer colors, consistent with an origin in the late type 
progenitor disk. The inner one (t2) is slightly redder, 
and this might be due to contamination  by the
population of the disk-like component  associated with  HCG 54a.
Therefore the morphology of HCG 54 can be well explained
by the collision of two similar mass disk systems. 
The model does not allow further test or comparison, 
since no prediction is given neither for the kinematics nor for the
stellar populations/colors of the colliding systems. 
Here we will try to shed more light in the
identification of  the involved systems based on the
rest of the information provided by our optical and HI data.

We have tried to determine the number of present galaxies using the
observed HI mass (log[M(HI)$_{\rm obs}$/M$_{\odot}$] = 8.75).
 We have calculated the  expected  mass as a function of 
optical luminosity and morphological type
 via the relationships obtained by Haynes \& Giovanelli (1984). 
Assuming that  HCG 54 is composed of 
4 irregular galaxies as classified in Hickson catalog, we predict a mass
of  log[M(HI)$_{\rm pred}$/M$_{\odot}$] = 
9.16 $\pm$ 0.23, but for a single irregular galaxy
the expected  mass would be  log[M(HI)$_{\rm pred}$/M$_{\odot}$] = 
9.19 $\pm$ 0.30. Unfortunately these cases are indistinguishable since they are within one sigma, and all 
we can conclude is that the HI content of 
HCG 54 does not deviate significantly from normal.

HCG 54a is the best defined object of the system, with a defined 
nucleus, embedded in a very elongated blue component and surrounded 
by a  rounder and redder stellar envelope.
This ensemble seems to be aligned with a perturbed but still visible
 HI velocity gradient. This might indicate that HCG 54a is one
 of the interacting galaxies, whose disk is a red round
 component that still keeps memory of the original rotation according to 
the HI kinematics (Fig. 7b).  We find
 however several arguments against this hypothesis.  The atomic gas is
 decoupled from the stellar component both in morphology and
 kinematics.
The optical emission is displaced to the north by $\sim$ 1 kpc with
respect to the atomic gas, and this shift is too large to be explained
only by stripping of the northern HI into the long tail.
Furthermore, the HI velocity gradient does not have a counterpart
in the optical velocities. The signature of
a rotation curve is not found along any of the observed directions, 
except for a small velocity gradient 
of 50\kms\ in the inner 5\arcs\ of HCG 54a (see Sect. 3.2).
The HI radial velocities along the direction of the optical spectra 
differ from the optical ones by $\sim$  30\kms, also consistent 
with a decoupling between the stellar and gaseous components.
Finally, since the HI velocity gradient is the only trace of a
possible disk system in the central part of HCG 54, then the second
disk would be a minor body when compared with HCG 54a, in
contradiction with the equal masses involved in the simulations.
Furthermore, if the observed HI velocity gradient of $\sim$ 180\kms\ had
to be attributed to a single galaxy, it would be more characteristic of
an Sb type system, which is not observed here.

A more plausible scenario is that we are observing 
a mix of  the two components of the original systems, that  have almost 
coalesced and  now present a similar distribution
as found in the simulations.  The  irregular
shape of the optical velocity curve supports an advanced merger stage.
HCG 54b shows a clear WR feature in the spectrum, indicative of 
a recent and  strong burst of star formation that was possibly triggered
by the interaction, while the underlying population, if present,
is negligible, as suggested by the very high equivalent
widths of the brightest emission lines (see Sect. 3.1.).
In fact we found a possible signature of gas infall in our spectra 
(see Sect. 3.2) that would be feeding the recent 
violent star formation observed.
HCG 54c and d do not have a well defined photometric
structure, but consist of an aggregate of knots,
 and have HII regions features in their spectra.  The very
blue color of HCG 54d suggests a pure star formation episode 
younger than several million years old.  They plausibly constitute the
debris of the interacting systems.
The HI component of the kinematics we observe might be the 
relic of the orbital motions of the interacting systems. However 
 projected radial motions cannot be negelected and 
detailed numerical simulations including a kinematical
study are needed to interpret the data.

In summary, no clear individual galaxies could be identified 
toward HCG 54 while strong signs of interactions and a
merger are found. The overall morphology is in good
agreement with the model by Hernquist \& Spergel (1992), and we
propose that HCG~54 is in the final stages of a merger of two
similar mass disk systems.

\subsection{Formation of the HI tidal tail}

While a chance projection is 
possible, the location of A1127+2054, a dwarf galaxy at the redshift
of the group, closely aligned with the large HI tidal tail is 
strongly suggestive of its role in shaping the appearance and
the extent of the HI tail.  A tidal tail pointing back at
the responsible intruder is a commonly observed generic feature, 
particularly soon after the closest approach, as the exchange of 
momentum between the intruder and the tidally disrupted material 
tends to bring them close to the common area in the phase space.

A1127+2054 is not likely to be a tidal dwarf since it is not
located within the tail and has its own rotating HI disk or cloud.
The fact that the associated HI feature is not well centered on 
the stellar body is consistent with the tidal disruption scenario.  
No stellar counterpart to the HI tail is detected down to a 
surface brightness level of 27 mag/(\arcs)$^2$ in R along its length.

This tidal disruption scenario has some potential difficulties however.
Assuming an inclination of $30^\circ$, the estimated dynamical
mass for A1127+2054 inside 2.1 kpc radius is only about $4\times 
10^8 M_\odot$, which is about 4\% of the dynamical mass of HCG~54.
On the other hand, the estimated HI mass {\it alone} for the 20 kpc 
long tidal tail is about $10^8 M_\odot$ (see Sect.~\ref{sec:HI}), 
and raising such a massive tidal tail from the gravitational 
potential of HCG~54 may require a far more massive perturber.

An alternative scenario for the formation of the large 
HI tail is the merger of two equal mass progenitors 
as we have already proposed in the previous section in order to
account for the disrupted stellar features around HCG~54. 
In numerical simulations of galaxy mergers including
gas particles such as by Mihos \& Hernquist (1996), formation of
one or more massive gaseous tails is commonly seen.  For example,
the projected appearance of the HI tail with respect to the
stellar remnant in HCG~54 is similar to the ``$t\sim 70$''
stages shown in Fig.~1 \& 2 of Mihos \& Hernquist (when 
mirrored about the vertical axis).  This
corresponds to a late stage in this particular simulation
where the stellar cores finally merge, similar to what we
infer in HCG~54.  If we are allowed to press this comparison
a little further, the extended HI structures seen to the south and 
southeast of the stellar body may be interpreted as the tidal
tail emerging from the second progenitor, as seen in the 
simulation.  Of course, these inferences are qualitative 
at best since the initial conditions of the Mihos \& Hernquist 
simulation are likely to be different from those of the
merger involved in HCG~54.  

In this scenario, an active role of 
A1127+2054 in shaping the appearance of the large HI tail
is still allowed.  For example, a dwarf companion UGC~957
is found within one of the large HI tails resulting from 
the merger of two massive disk galaxies in NGC~520 (see
Hibbard \& van Gorkom 1996) even though UGC~957 may be only
a third party to the ongoing or recent merger.  

\section{Conclusions}

We propose that HCG~54 is a system that consists of at least
two late type galaxies plus a dwarf companion
involved in recent tidal interactions and
a merger.  Two low mass disk systems merged and induced the formation 
of a central blue, bar-like structure embedded in a red, round 
stellar envelope, outside which lies
several stellar shells. A luminous starburst is associated with 
HCG~54b while HCG 54c and d are the remnant debris of the recent
merger undergoing modest star formation activity. 
This merger is probably in a very advanced state.  The 20 kpc
long gaseous tidal tail discovered in HI may be a result of
the same merger or driven by the dwarf galaxy A1127+2054 found 
at a projected distance of 8 kpc from the tip.  In either 
scenario, it is likely playing a role in shaping the appearance of 
this tidal tail.  Unfortunately current theoretical or numerical 
models do not provide
information on the kinematics or the distribution of induced
star formation in low mass disk mergers.  Few similar systems
have been discussed in the literature, and 
the present data should provide valuable inputs for future
modeling of mergers involving low mass galaxies.

\vfill\eject

\parindent 0pt

\begin{figure*}
\vskip 2truecm
\hfill
\caption{Full field of the R-Johnson band optical image of HCG 54
obtained as explained 
in the text (Sect. 2.1). The dotted lines delineate the HI integrated emission 
at a level of 0.5 $\times$ 10$^{20}$ atoms cm$^{-2}$. We have labeled the 
three dwarf galaxies discussed in the text (Sect. 3.2).}
\end{figure*}

\begin{figure*}
\hfill
\caption{Combined frame of B, R and $r'$ bands in order to 
better recover the faintest emission. Contours are shown for the central
part that appears saturated in our greyscale. The objects identified
 by Hickson et al. (1989) as the members of HCG 54 are indicated with letters
a to d. We also mark the tidal optical features described in Sect. 3.1
t1 to t4, and an example of the unresolved knots (k) that we detect
close to HCG 54.}
\label{fig:B+R}
\end{figure*}

\begin{figure*}
\hfill
\caption{B-R map of HCG 54. Averaged
colors of different regions are indicated by arrows, and the displayed range 
goes from 0 (white) to 0.8 (black). We have 
superimposed two R band isophotes as a reference. }
\end{figure*}

\begin{figure*}
\hfill
\caption{R band image of HCG 54 showing the direction of the slits 
used for the spectroscopical observations in the main body of
 HCG 54 (see Sect. 2.2), and the sections 
indicated in Tables 3 and 4. }
\label{fig:Rimage}
\end{figure*}
 
\begin{figure*}
\hfill
\caption{(a) Up: Spectrum in the direction joining HCG 54a and b (sp3), where
the position is given in offsets 
with respect to the peak of emission of HCG 54b. The indicated regions
are described in Sect. 3.2. Bottom: R band profile along the 
slit.  }
\label{fig:spec54ab}
\end{figure*}

\setcounter{figure}{4} 

\begin{figure*}
\hfill
\caption{(b) Up: Spectrum in the direction joining HCG 54c and d (sp4), where
the position is given in offsets with respect to the peak of emission 
of HCG 54c. Bottom:  R band profile along the slit.  }
\label{fig:spec54cd}
\end{figure*}
 
 \setcounter{figure}{4}

\begin{figure*}
\hfill
\caption{(c) Spectrum in the direction joining HCG 54d and A1126+2051 
(sp1, Fig. 4),
for the sections where emission was detected.
Positions are given in
offsets with respect to the continuum peak  of HCG 54d along
the slit.}
\label{fig:MHISD}
\end{figure*}

\begin{figure*}
\hfill
\caption{Full  spectrum of HCG 54a. }
\label{fig:spec2}
\end{figure*}

\begin{figure*}
\hfill
\caption{(a) Full  spectrum of HCG 54b.}
\label{fig:spec1a}
\end{figure*}
 
\setcounter{figure}{6}

\begin{figure*}
\hfill
\caption{(b) Selected range of  the same spectrum shown in (a).}
\label{fig:spec1b}
\end{figure*}

\begin{figure*}
\hfill
\caption{(a) Left: HI column density contours 0.5, 0.7, 1.2, 2.0, 
2.7, 3.4, 4.4, 6.7, 8.4, 10.1, 13.5, 16.9, 20.2 to 22.6 
$\times 10^{20}$ atoms cm$^{-2}$ overlapped on the R image of HCG 54.
The atomic emission has been integrated in the velocity
range 1334 - 1490 \kms .     The synthesized beam is
20\arcs  $\times$ 16\arcs . 
(b)  Right: Map of the first--order moment of the HI radial velocity field.   
The scale goes as in the wedge where the numbers indicate heliocentric
velocities in km s$^{-1}$. The contours go from 1340 to 1490 \kms\ with a
step of 10 \kms. The beam size is 20\arcs  $\times$ 16\arcs .}
\label{fig:HImoments}
\end{figure*}
 
\begin{figure*}
\hfill
\caption{Channel maps of the 21 cm line radiation 
superimposed on the R image of the group.
Heliocentric velocities are
indicated in each panel. Contours correspond to
2.4, 4.0, 5.6, 7.3, 8.9, 10.5, 12.1, 13.7, 15.3 K, and the
rms noise of the maps is 1.2 K.  The synthesized beam
(15\arcsper 8 $\times$ 14\arcsper 5)
is plotted in the upper  left panel.    }
\label{fig:HICH1}
\end{figure*}
 
\setcounter{figure}{8} 

\begin{figure*}
\hfill
\caption{(Cont).   }
\label{fig:HICH2}
\end{figure*}

\begin{figure*}
\hfill
\caption{Integrated HI emission as a function of the 
heliocentric velocity for the range where we detect emission. The solid line
corresponds to our VLA data, while the dashed ones are single dish data
from Huchtmeier (1997).}
\label{fig:HIspec}
\end{figure*}

\begin{figure*}
\hfill
\caption{(a) Upper-left: R band emission of A1127+2054 in greyscale with 
overlapped isophotes 22.6, 22.9, 23.3, 24.1 mag arcsec$^{-2}$. 
(b) Upper right: Light profile of this galaxy in the R band 
as a function of the equivalent radius.
(c) Lower left: Integrated HI emission of this galaxy where the contours correspond to
3.3, 6.6, 10.0 and 11.6 $\times$ 
10$^{20}$ atoms cm$^{-2}$, and overlapped on the R image. 
(d) Lower-right: 
Map of the first--order moment of the HI radial velocity field.   
The scale goes as in the wedge where the numbers indicate heliocentric
velocities in km s$^{-1}$.  }
\label{fig:DW}
\end{figure*}

\begin{acknowledgements}
 We wish to thank Dr. Masegosa and Dr. M\'arquez for the help with the R
observations at the NOT and their valuable comments.
 LV--M, AO and JP are
partially supported by DGI (Spain)
Grant AYA2000-1564 and Junta de Andaluc\'{\i}a (Spain).
JMV is partially supported by DGI (Spain) Grant AYA2001-3939-C03-01 
and Junta de Andaluc\'{\i}a (Spain).

\end{acknowledgements}
\vfill\eject

\clearpage
                                  
\end{document}